# Anomalous Charge Density Wave State Evolution and Dome-like Superconductivity in $CuIr_2Te_{4-x}Se_x$ Chalcogenides

**Mebrouka Boubeche[1], Ningning Wang[2], Jianping Sun[2], Pengtao Yang [2], Lingyong Zeng[1], Qizhi Li[3] Yiyi He[1], Shaojuan Luo[4], Jinguang Cheng[2], Yingying Peng[3], Huixia Luo[1]\***

[1] School of Materials Science and Engineering, State Key Laboratory of Optoelectronic Materials and Technologies, Key Lab of Polymer Composite & Functional Materials, Sun Yat-Sen University, No. 135, Xingang Xi Road, Guangzhou, 510275, P. R. China

[2] Beijing National Laboratory for Condensed Matter Physics and Institute of Physics, Chinese Academy of Sciences and School of Physical Sciences, University of Chinese Academy of Sciences, Beijing 100190, China

[3] International Center for Quantum Materials, School of Physics, Peking University, Beijing 100871, China

[4] School of Chemical Engineering and Light Industry, Guangdong University of Technology, Guangzhou, 510006, P. R. China

E-mail: luohx7@mail.sysu.edu.cn

**Abstract**

We report the anomalous charge density wave (CDW) state evolution and dome-like superconductivity (SC) in $CuIr_2Te_{4-x}Se_x$ ($0 \leq x \leq 0.5$) series. Room temperature powder X ray-diffraction (PXRD) results indicate that $CuIr_2Te_{4-x}Se_x$ ($0 \leq x \leq 0.5$) compounds retain the same structure as the host $CuIr_2Te_4$ and the unit cell constants $a$ and $c$ manifest a linear decline with increasing Se content. Magnetization, resistivity and heat capacity results suggest that superconducting transition temperature ($T_c$) exhibits a weak dome-like variation as substituting Te by Se with the maximum $T_c = 2.83$ K for $x = 0.1$ followed by suppression in $T_c$ and simultaneous decrease of the superconducting volume fraction. Unexpectedly, the CDW-like transition ($T_{CDW}$) is suppressed with lower Se doping ($0.025 \leq x \leq 0.2$) but re-emerges at higher doping ($0.25 \leq x \leq 0.5$). Meanwhile, the temperature-dependent XRD measurements show that the trigonal structure is stable at 20 K, 100 K and 300 K for the host sample and the doping composition with $x = 0.5$, thus we propose that the behaviour CDW-like transition arises from the disorder effect created by chemical doping and is not related to structural transition. The lower and the upper critical fields of these compounds are also addressed.

# 1. Introduction

Transition-metal chalcogenides (TMCs) own plentiful structural chemistry and physical properties such as nontrivial topological properties [1-3], charge density wave (CDW) [4-6], superconductivity (SC) [7-10], Mott transition [11], extremely large magnetoresistance [12,13], spin glass [14], and so on. Specifically, the coexistence/competition of SC and CDW is a long history but still a hot topic in condensed matter physics. CDW and SC are two fully distinct cooperative states, both of which happen because of the instabilities of Fermi surface (FS) that result in the splitting of FS and the density of states (DOS) at FS surface decrease below their corresponding transition temperatures. So far, abundant results demonstrate that a superconducting dome can usually be seen on the edge of a CDW/structural instability by the application of hydrostatic pressure [15,16], chemical doping [17-22], or gating [23]. Among chemical doping methods, the isovalent substitution is a commonly used strategy to study the key relation between SC and instabilities in some other order parameters. A recent, prominent case is $1T\text{-}TaS_{2-x}Se_x$, in which SC occurs when the commensurate CDW (CCDW) Mott phase vanishes, and it coexists with the nearly commensurate (NCCDW) state [24]. Moreover, the isoelectronic S substitution for Se in $2H\text{-}TaSe_2$ leads to the emergence of a robust superconducting order in the $2H\text{-}TaSe_{2-x}S_x$ ($0 \le x \le 2$) series. In the case of $2H\text{-}TaSe_{2-x}S_x$ ($0 \le x \le 2$) compounds, the CDW is suppressed and the SC is maximized with crystallographic disorder, and the superconducting transition temperature ($T_c$) of the doped compound is surprisingly higher than those of two end compounds [25]. Another example of isoelectronic substitution material is $TaSe_{2-x}Te_x$ ($0 \le x \le 2$). The isoelectronic Te replacement for Se in $2H\text{-}TaSe_2$ not only leads to two polytypes and two polymorphs, but also enhances the $T_c$. The $T_c$s of the 1T and 3R polymorph $TaSe_{2-x}Te_x$ are both higher than that of the undoped $2H\text{-}TaSe_2$. Especially, the maximum $T_c$ of $3R\text{-}TaSe_{2-x}Te_x$ variants is five times larger than that of the undoped $2H\text{-}TaSe_2$ sample [26]. Moreover, the $3R\text{-}TaSe_{2-x}Te_x$ variants exhibit the coexistence of SC and CDW, with $3R\text{-}TaSe_{1.9}Te_{0.1}$ initially displaying an incommensurate (ICCDW) and then transitioning to a CCDW phase upon further cooling. This phenomenon likely emerges either from an unexpected influence of the layer stacking order on the electronic properties, which is likely to be dominated by the characteristics of the single layer or from exceptional dependence of $T_c$ on the subtle differences in the properties of the single layers in this type of compounds [26]. Besides, SC is emerged or enhanced upon isoelectronic Se substitution for Te/S in the parent transition-metal chalcogenides (e.g. $1T\text{-}PdTe_2$, $1T\text{-}TaS_2$, $ZrTe_3$) [27-30]. The enhancement of $T_c$ in $1T\text{-}PdSeTe$ is considered to be arising from the possible existence of metallic Pd or PdTe stripes and/or some dedicated structural disorder caused by a rather small amount of Se deficiency. Except for the transition-metal chalcogenide compounds, Se chemical doping has been confirmed to be an effective tool to tune the CDW and SC in other systems like $Eu_3Bi_2S_4F_4$ [31], $Nb_2Pd(S_{1-x}Se_x)_5$ [32] and so on.

Very recently, the occurrence of SC in a CDW materials has been observed in the quasi-two-dimensional $Cu_{0.5}IrTe_2$ ($CuIr_2Te_4$) chalcogenides with a NiAs defected trigonal structure (space group $P3$-$m1$) as presented in the inset of **Fig. 1(a)**, in which Cu is intercalated between the Te-Ir-Te layers [33]. The $T_c$ is around 2.5 K, and the CDW transition temperature ($T_{CDW}$) is about 250 K (from heating) and 186 K (from cooling). First-principles calculations further reveal that the electronic DOS in the vicinity of the Fermi energy fundamentally descend from the Ir $d$ and Te $p$ orbitals [33]. Therefore, it is reasonable to tune the SC and CDW in the parent $CuIr_2Te_4$ telluride chalcogenide with isoelectronic Se substitution for Te-site by controlling the Fermi energy $E_F$ or creating disorder.

On the other hand, copper-based chalcogenide spinels display rich physical properties such as magnetic ordering [34-36], metal-insulator transition (MIT) [37-39], and SC [8, 40-42]. It is well known that the sulpo-$CuRh_2S_4$, where the Cu atoms occupy the tetrahedra sites and the Rh atoms occupy the octahedra sites, shows SC with $T_c = 4.70$ K [31]. Another spinel $CuIr_2S_4$ displays temperature-induced MIT near 226 K with structural phase transition, displaying hysteresis on cooling and heating [39]. Selenospinel $CuIr_2Se_4$ (isostructural to $CuIr_2S_4$), however, remains ordinary metal above 0.5 K at ambient pressure but shows a MIT above 2.8 GPa [43,44]. Despite the absence of metal-metal pairing or charge ordering in $CuIr_2Se_4$, it seems to be at the margin of such performance [45], i.e., it possibly has a nascent propensity to such fluctuation due to the robust spin-orbit coupling of $5d$ Ir and the geometrical frustration inherent to the spinel structure. Indeed, SC in $CuIr_2Se_4$ spinel has been prompted by Pt substitution for Ir [8]. Therefore, these differences in structural and electronic properties between the layered $CuIr_2Te_4$ and the spinel $CuIr_2Se_4$ have attracted our attention as a starting point to explore the effect of the isoelectronic Se substitution for Te site on the SC and CDW in $CuIr_2Te_4$ compound and whether it can facilitate SC near a CDW state.

For this matter, we successfully synthesized and systemically investigated the crystallographic and physical properties of the polycrystalline $CuIr_2Te_{4-x}Se_x$ ($0 \leq x \leq 0.5$) series. Yet, several attempts have been failed to produce single crystal for this family. The analysis of XRD measurements under several temperatures (20 K, 100 K and 300 K) indicates that $CuIr_2Te_{4-x}Se_x$ ($0 \leq x \leq 0.5$) samples maintain the same layered structure as the $CuIr_2Te_4$, but mixed layered phase and cubic phase can be found for $x \geq 0.50$ even near the end. Resistivity and magnetic susceptibility measurements both show consistently that isoelectronic Se substitution for Te in $CuIr_2Te_4$ first favours the SC near a CDW accompanied with the anomalous CDW state evolution in two sides of dome-like SC. The optimal doping composition is $CuIr_2Te_{0.9}Se_{0.1}$ with the highest $T_c = 2.83$ K. The CDW-like states disappear at lower doping content but re-emerge at higher Se doping concentration.

## 2. Methods

Polycrystalline $CuIr_2Te_{4-x}Se_x$ ($0 \le x \le 0.5$) compounds are made by means of a vacuum shield solid-state reaction method. Cu, Ir, Te and Se powders were weighted in the stoichiometric ratios and ground with a mortar. Then the mixture was sealed in the evacuated quartz tubes and sintered at 1023 K for five days. The as-prepared precursors were then reground, compressed into pellets and heated at 1023 K for ten days. The polycrystalline $CuIr_2Te_{4-x}Se_x$ samples are not air-sensitive.

To determine phase purity and analyze the detailed crystal structure, powder X-ray diffraction (PXRD) characterization was carried out at room temperature by using MiniFlex, Rigaku apparatus with Cu Kα1 radiation. Rietveld model in FULLPROF suite software was used to perform the structural parameter refinements for the obtained polycrystalline $CuIr_2Te_{4-x}Se_x$ compounds. Besides, XRD measurements under 20 K, 100 K and 300 K were carried using a Mo K (17.4 keV) microspot X-ray source and a Pilatus 2D detector. All temperature evolution measurements are conducted through warming process. The elements ratios and their distributions were examined via scanning electronic microscope (SEM) EM-30AX PLUS from Kurashiki Kako Co. Ltd, Japan, equipped by an energy dispersive X-ray spectroscopy (EDXS) detector. The temperature-dependent electrical resistivity ($\rho(T)$) of all the samples and the specific heat capacity ($C_p(T)$) measurements we carried out using a quantum design physical property measurement system (PPMS). The DC magnetic susceptibilities were measured using a superconducting quantum interference device (SQUID) magnetic property measurement system (MPMS) under 10 Oe. Using resistivity $\rho(T)$ data, $T_c$s are extracted from the midpoint of the drop region, and the extrapolations of the abrupt slope of the susceptibility and the normal state susceptibility, $T_c$ from the specific heat capacity ($C_p(T)$) is estimated by the equal area entropy construction method. $T_{CDW}$ values have been estimated from the minimum of the temperature-dependent derivative of resistivity ($d\rho/dT(T)$) and the maximum of the temperature-dependent derivative of magnetic susceptibility ($d\chi/dT$) measured under 10 kOe.

## 3. Results and discussions

We performed detailed structural investigation using the Rietveld refinement technique to check the phase purity and crystal structure of all the synthesized polycrystalline $CuIr_2Te_{4-x}Se_x$ ($0 \le x \le 0.5$) series. The refinement for the representative $CuIr_2Te_{3.9}Se_{0.1}$ compound is shown in **Fig. 1(a)**. An analysis of XRD patterns shows that $CuIr_2Te_{3.9}Se_{0.1}$ compound has a trigonal structure with lattice contents $a = b = 3.9382$ Å, $c = 5.3952$ Å, which is slightly lower than those of the parent compound $CuIr_2Te_4$ with $a = b = 3.9397$ Å, $c = 5.3964$ Å [32]. The Rietveld refinements for the other investigated polycrystalline samples are depicted in the supplemental information (**Fig. S1**). Besides, the refinement results confirm that all these $CuIr_2Te_{4-x}Se_x$ ($0 \le x \le 0.5$) series preserve the basic trigonal structure with the space group *P-3m1* as presented in the inset of **Fig. 1(a)**, though tiny unreacted Ir appears in all the specimens, where Cu atoms occupy the octahedral sites $1b$ (0 0 ½), Ir atoms occupy the sites $1a$ (0 0 0), Te/Se atoms are in sites $2d$ (⅓ ⅔ z). Room temperature XRD patterns of the polycrystalline $CuIr_2Te_{4-x}Se_x$ ($0 < x < 0.5$) compounds are shown in **Fig. 1(b)**. The variation of the (002) diffraction peaks for different Se concentrations is revealed in the right side of **Fig. 1(b)**, where all the (002) peaks shift towards higher angle with increasing selenium concentration in $CuIr_2Te_{4-x}Se_x$ ($0 \leq x \leq 0.5$). Indeed, as depicted in **Fig. 1(c)**, the lattice constants $a$ and $c$ both manifest linear decrease with the increase of Se concentration $x$ for our synthesized $CuIr_2Te_{4-x}Se_x$, which is in good agreement with Vegard's law [46]. The reason behind this is the smaller ionic radii of Se (1.98 Å) compared with that of Te (2.21 Å) [47]. Hence, a consistent reduction of the unit cell volumes of the $CuIr_2Te_{4-x}Se_x$ compounds is expected. Nevertheless, samples with higher Se content (the vicinity of the spinel $CuIr_2Se_4$) have not been successfully synthesized at present. In addition, we further used SEM-EDX to examine the ratio and the distribution of the elements and found that the experimental percentages of Cu:Ir:Te:Se for $CuIr_2Te_{4-x}Se_x$ samples are close to the initial percentages (see **Fig. S2**). As shown in **Fig. S3**, we can see a homogeneous distribution for all the elements in these $CuIr_2Te_{4-x}Se_x$ polycrystalline samples.

The raw resistivity data as a function of temperature is plotted in **Fig. S4**. The temperature-dependent normalized resistivity ($\rho/\rho_{300K}$) of $CuIr_2Te_{4-x}Se_x$ ($0 \leq x \leq 0.5$) polycrystalline series under zero magnetic field is given in **Fig. 2a**. The specimens demonstrate a metallic behaviour in the temperature region of 3 - 300 K, but there is abnormal hump in some samples. A sharp drop of resistivity can be seen at low temperatures, identifying the onset of SC. Normalized resistivity $\rho/\rho_{300K}$ near $T_c$ is shown in **Fig. 2b**. $T_c$ gradually increases up to 2.83 K for the optimal $CuIr_2Te_{3.9}Se_{0.1}$. From **Table 1** and **Fig. S5**, we can see that the increase of $T_c$ is followed in an increase residual resistivity ratio (RRR = $R_{300K}/R_{8K}$) from 4.16 for the undoped sample to 5.88 for the doped sample with the highest $T_c$ ($x = 0.1$). Besides, from **Fig. 2a**, we can see that the anomalies hump disappears at very low Se doping concentration region, indicating the suppression of CDW-like transition temperatures ($T_{CDW}$), which is determined by the minimum of $d\rho/dT$ (*see* inset of **Fig. 2(a)**). Meanwhile, the Se doped samples in the region of $0 < x \leq 0.3$ exhibit sharp superconducting transition, implying that the samples is highly homogeneous (see **Fig. 2b** and **Table 1**). Unexpectedly, the hump anomalies associated with CDW-like transition in the $\rho(T)$ curves re-emerges when $x = 0.25$. For $0.25 \leq x \leq 0.5$, the $T_{CDW}$ transition anomalies steadily increases with increasing Se doping, while the RRR (RRR = $R_{300K}/R_{8K}$) sharply decreases from 3.02 for $x = 0.25$ to 1.87 for $x = 0.5$ **(see Table 1 and Fig. S5)**. The decrease of RRR suggests that Se doping induces disorder significantly and Se ions are effective scattering centres [48-50], which could account for the reemergence of the CDW.

The magnetic susceptibility measurement provides substantial evidence for bulk SC. The zero-field cooling (ZFC) temperature dependence of the normalized magnetization ($4\pi\chi$) data under 10 Oe for $CuIr_2Te_{4-x}Se_x$ samples are given in **Fig. 2c and Fig. S5**, which display a diamagnetic behaviour below $T_c$s. The change in $T_c$s is found to be in good agreement with the electrical transport data, whereas the superconducting volume fraction is getting smaller with higher doping content. In particular, the superconducting volume fraction for those samples with the re-emergence of CDW-like transition for $x \geq 0.25$ is much smaller than those of other samples without CDW-like transition, suggesting the competition between SC and CDW-like transition. We further performed the magnetization hysteresis under 10 kOe from cooling and heating process as shown in **Fig. 2d and Fig. S5**, these compounds with $x \geq 0.25$ also exhibit the CDW-like related magnetic anomalies. The inset in **Fig. 2d** represents the cooling temperature differentiated magnetization ($d\chi/dT(T)$) curves that have been used to define the values of $T_{CDW}$. These values are in good agreement with those data derived from $\rho(T)$ measurements.

With the purpose of revealing whether the abnormal hump is related to a structural phase transition, we have performed XRD measurements on the parent $CuIr_2Te_4$ and highest doping sample $CuIr_2Te_{3.5}Se_{0.5}$ at 20 K, 100 K and 300 K. From **Fig. 3**, we can see that the three XRD patterns measured at different temperature are similar, which can be successfully indexed to the space group *P3-m1*. Thus, the abnormal hump is likely not linked to the structural transition. It indicates that there is no structural phase transition at low temperature, which differs from the case for $IrTe_2$ where the resistivity anomaly has been proved associated to the structural phase transition from trigonal to monoclinic structure at cooling [18, 20, 51]. However, there is no strong CDW signal can be detected, which may be ascribed to the polycrystalline nature of our samples. We have also realized that it will be more clearly if the observed abnormal humps appear in the crystalline samples instead of the polycrystalline compounds. However, we have tried a few methods including self-flux method, vapor transport method and modified Bridgman method to grow the crystalline samples, but all failed to obtain the target crystalline samples so far.

To further confirm the bulk SC in the $CuIr_2Te_{4-x}Se_x$ system, the optimal doping polycrystalline $CuIr_2Te_{3.9}Se_{0.1}$ composition is characterized by low-temperature specific heat capacity ($C_p$) measurement. The main panel of **Fig. 4a** displays the plots of $C_p/T$ against $T^2$ for 0 and 10 kOe applied fields. A clear sharp anomaly in zero magnetic field is observed around 2.81 K as presented in the inset **Fig. 4(a)**, which is close to that determined by the resistivity and susceptibility. Besides, the specific heat jump is suppressed upon the application of the magnetic field 10 kOe, indicating that the upper critical field for $CuIr_2Te_{3.9}Se_{0.1}$ is less than 10 kOe. The zero-field specific heat above $T_c$ can be well fitted to $C_p/T = \gamma + \beta T^2$ (see the dashed red line in **Fig. 4a**), where $\gamma$ is the constant of electronic contribution to the specific heat ($C_{el}$) and $\beta$ is the lattice constants in the second term of the phonon contribution ($C_{ph.}$). $\gamma$ value is about 10.84 mJ mol$^{-1}$ K$^{-2}$, and $\beta$ value is near 3.51 mJ mol$^{-1}$ K$^{-4}$. Further, the Debye temperature ($\Theta_D$) can be calculated from the formula $\Theta_D = (12\pi^4 nR / 5\beta)^{1/3}$, where n = 7 is the number of atoms per formula unit and R is the gas constant. $\Theta_D$ value for the CuIr$_2$Te$_{3.9}$Se$_{0.1}$ sample is 157 K. Hereafter, we calculated the electron-phonon coupling constant ($\lambda_{ep}$) using the value of $\Theta_D$ and $T_c$ values from the inverted McMillan equation [52]: $\lambda_{ep} = \frac{1.04 + \mu^*\ln(\Theta_D / 1.45T_c)}{((1-1.62\mu^*)\ln(\Theta_D / 1.45T_c) - 1.04}$ , where $\mu^* = 0.15$ is the repulsive screened coulomb parameter. The value of $\lambda_{ep}$ is estimated to be 0.65. We then calculated the DOS at the Fermi level $N(E_F)$ from the formula $N(E_F) = 3\gamma / (\pi^2 k_B^2 (1 + \lambda_{ep})$ , where $k_B$ is Boltzmann's constant. $N(E_F)$ is calculated to be 3.11 states/eV/f.u. (f.u. = formula unit), which is little higher then that of the pristine CuIr$_2$Te$_4$. The enhancement in the superconductivity in CuIr$_2$Te$_{4-x}$Se$_x$ compounds can be explained according to the enhancement of the electron-phonon coupling ($\lambda_{ep}$) and the increase of DOS of Fermi surface $N(E_F)$ by Se substitution as compared to the host CuIr$_2$Te$_4$. We calculated the specific heat jump $\Delta C$ ($T = T_c$) as the difference between the $C_{el}$ at $T_c$ and the normal state (illustrated by the solid red line in the inset of **Fig. 4(a)**. The normalized specific heat jump value $\Delta C_{el}/\gamma T_c$ is estimated to be 1.51, which is slightly higher the Bardeen-Cooper-Schrieffer (BCS) weak-coupling limit (1.43), which confirm the bulk nature of SC in CuIr$_2$Te$_{3.9}$Se$_{0.1}$.

Since the lower critical and the upper critical fields ($\mu_0 H_{c1}(0)$ and $\mu_0 H_{c2}(0)$) are essential properties of superconductors, we next estimate the $\mu_0 H_{c1}(0)$ and $\mu_0 H_{c2}(0)$. As represented in **Fig. 4(b)**, $\mu_0 H_{c1}(T)$ for the optimal compound CuIr$_2$Te$_{3.9}$Se$_{0.1}$ has been estimated from the magnetization isotherms in the temperature range 1.8 K - 2.8 K (see lower inset of **Fig. 4(b)**)**.** The profile of the $M(H)$ curves points that CuIr$_2$Te$_{3.9}$Se$_{0.1}$ is a type-II superconductor as evidenced by the linear shielding ("Meissner line") at low fields (see solid straight line in inset). Above ~ 300 Oe, the shielding reduces as magnetic flux starts to penetrate the bulk and the system gets in the vortex state. To figure out the $\mu_0 H_{c1}(0)$, we pursued the method which has been used before for different SC as demonstrated in the insets of **Fig. 4(b)**. The demagnetization effect has been considered to get an accurate value of $\mu_0 H_{c1}(0)$. We can get the demagnetization factor ($N$) value from the formula $N = 4\pi\chi_V + 1$ , where $\chi_V = dM/dH$ is the slope of linear fitting (see the upper inset in **Fig. 4(b)**). The estimated value of $N$ is 0.56. As shown in the lower inset of **Fig. 4(b)**, the fitted purple linear line describes the Meissner shielding effects at low fields based on the formula $M_{\text{Fit}} = fH + e$ (Meisner line) in the region of low magnetic fields, where $f$ represents the slope of the linearly fitted $M(H)$ data and $e$ is the intercept. We can subtract Meissner line from the magnetization $M$ for each isotherm ($M$-$M_{Fit}$ ($H$)) to determine the value of $\mu_0 H_{c1}*$. As depicted in the top inset in **Fig. 4(b)**, $\mu_0 H_{c1}*$ values are determined from 1 % $M$ at the field when diverges below the fitted data ($M_{Fit}$). By considering $N$, we can obtain $\mu_0 H_{c1}(0)$ from the expression: $\mu_0 H_{c1}(T) = \mu_0 H_{c1}^*(T)/(1-N)$. We can further calculate the $\mu_0 H_{c1}(T)$ data using the formula: $\mu_0 H_{c1}(T) = \mu_0 H_{c1}(0)(1-\left(\frac{T}{T_c}\right)^2)$. The extrapolation of the $\mu_0 H_{c1}(T)$ data down to T = 0 K yields the value $\mu_0 H_{c1}(0) = 66$ mT for CuIr$_2$Te$_{3.9}$Se$_{0.1}$, which is almost 2.5 times larger than $\mu_0 H_{c1}(0)$ of the parent CuIr$_2$Te$_4$ (28 mT, see **Table 2**).

The upper critical $\mu_0 H_{c2}(0)$ for the CuIr$_2$Te$_{3.9}$Se$_{0.1}$ and CuIr$_2$Te$_{3.8}$Se$_{0.2}$ superconducting samples are estimated from low temperature $\rho(T)$ data under different applied fields (between 0 and 500 Oe). The temperature-dependent resistivity is presented in **Fig. 4(c-d)**. The resistivity transition gradually shifts to lower temperatures but seems to be a least susceptible to the applied magnetic field, indicating robust SC. The $T_c$ at each magnetic field is extracted and plotted in the inset of **Fig. 4(c-d)**. The derived $\mu_0 H_{c2}(T)$ diagrams of CuIr$_2$Te$_{3.9}$Se$_{0.1}$ and CuIr$_2$Te$_{3.8}$Se$_{0.2}$ are plotted in **Fig. 4(c-d)**, respectively. The values of $\mu_0 H_{c2}(0)$ are determined through Ginzberg-Landau (GL) and Werthamer-Helfand-Hohenburg (WHH) theories. We calculated $\mu_0 H_{c2}(0)$ values from the criteria 90 %, 50 % and 10 % of superconducting transition in resistivity data ($\rho_N$) using the GL equation: [53] $\mu_0 H_{c2}(T) = \mu_0 H_{c2}(0) * \frac{1-(T/T_c)^2}{1+(T/T_c)^2}$,. The $\mu_0 H_{c2}(0)$ values from 50 % $\rho_N$ criteria are 148 and 127 mT, respectively. On the other hand, the $\mu_0 H_{c2}$ values calculated from the simplified WHH equation for the dirty-limit SC: $\mu_0 H_{c2}(0) = -0.693*T_c \setminus (dH_{c2}/dT)\big|_{T_c}$ [54,55] where $(dH_{c2}/dT_c)$ denotes the slope of $\mu_0 H_{c2}(T)$ near $T_c$ (the color solid line). The obtained values of $\mu_0 H_{c2}(0)$ from WHH model for CuIr$_2$Te$_{4-x}$Se$_x$ ($x = 0.1$ and 0.2) from the 50 % $\rho_N$ criteria are 144 and 125 mT, respectively. The $\mu_0 H_{c2}(0)$ do not surpass the Pauli limiting field for the weak-coupling BCS superconductors $H^P = 1.86*T_c$ [56]. Therefore, the values of $H^P$ are estimated to 5.26 and 5.04 T, respectively. The Ginzburg-Landau coherence length ($\xi_{GL}(0)$) is extracted from this equation $H_{c2} = \phi_0/(2\pi\xi_{GL}^2)$ [56] at the 50 % criteria $\mu_0 H_{c2}(0)$, where $\phi_0 = 2.07 \times 10^{-3}$ T $\mu$m$^2$ is the flux quantum. $\xi_{GL}(0)$ values for CuIr$_2$Te$_{3.9}$Se$_{0.1}$ and CuIr$_2$Te$_{3.8}$Se$_{0.2}$ are 47.831 and 51.34 nm, respectively. **Table 2** gives a sight about the physical properties of our present studied compounds as compared to other previous reported telluride chalcogenides.

**Fig. 5** depicts the $T(x)$ phase diagram of $T_{CDW}$ and $T_c$ versus Se content ($x$) for CuIr$_2$Te$_{4-x}$Se$_x$, showing CDW-like order, metallic state and superconducting phase edges. In the beginning, with increasing Se content $x$, the CDW-like order has vanished around $x = 0.025$. The suppression of CDW-like transition first causes the enhancement of $T_c$ and $T_c$ reach to the highest value of 2.83 K at $x = 0.1$. Nevertheless, with a further increase of Se, the $T_c$ decreases and gives rise to a weak SC dome-like phase diagram. Unexpectedly, the higher Se concentration ($x > 0.2$) induces the reappearance of CDW-like state with a lower transition value than that of the Se-free host material. Such tendency of CDW has been found by iodine doping for Te [57] but not by Zn doping in Cu site or Ru, Al, Ti doping in Ir site [58-60]. Thus, the CDW-like transition in this system seems to be dopant-dependent. Similar behaviour has also been reported for Tl-intercalated $Nb_3Te_4$ single crystals [61], which is ascribed to the disorder in the quasi 1-D Nb chains. Besides, $M_x TiSe_2$ where $M$ is $3d$ transition metal ($M$ = Mn, Cr, Fe) systems also manifested analogous phenomena where CDW was first suppressed and then re-appeared with higher intercalation concentration, which is due to the deformation degree of Se-Ti-Se sandwiches [62,63]. The re-occurrence of CDW is reported for $1T$-$TaS_{2-x}Se_x$ single crystals as well [64]. One additional case is $2H$-$TaSe_{2-x}S_x$ ($0 \leq x \leq 2$), where a large dome-like superconducting phase diagram is accompanied by the appearance of CDW at two ends [25], where disorder played a significant role in the behavior of CDW and SC. On the other hand, RRR has been widely considered to be an indication of disorder a dirty superconductor [48-50]. It is also well known that RRR ratio is reduced in the dirty-band case [65-68]. However, in our case, RRR ratio increases with increasing doping content in the lower doping region from 0 to 0.1, while decreases as $x$ arises in the higher doping region of 0.1 to 0.5. From **Fig. 5**, $T_c$ increases as the CDW transition is suppressed and completely disappears, while the $T_c$ decreases when the CDW transition reemerges and $T_{CDW}$ increased with the doping content. Overall, this system exhibits competing tendencies towards CDW and SC orders. Therefore, we suggest that the initial enhancement of the $T_c$ and disappearance of the CDW is due to the increase of density of states (DOS) at Fermi surface and the reemergence of CDW in over Se-doped region might result from the disorder scattering of Se-impurities and the decrease of DOS at Fermi surface because some portions of Fermi surface are removed by CDW gaping and consequently degrading superconducting $T_c$ [68]. Yet, further theoretical and/or experimental studies need to prove it.

## 4. Conclusion

In summary, our results highlighted the role of Se doping in superconductivity and CDW-like states on $CuIr_2Te_{4-x}Se_x$. We have successfully synthesized a series of $CuIr_2Te_{4-x}Se_x$ ($0 \leq x \leq 0.5$) polycrystalline compounds. Powder X-ray diffraction results from different temperatures demonstrate that the trigonal phase is stable at 20, 100 K and 300 K, indicating there is no structural phase transition at low temperature. By the means of the electrical resistivity, magnetic susceptibility and specific heat measurements of $CuIr_2Te_{4-x}Se_x$ ($0 \leq x \leq 0.5$) compounds, it can be found that when Se substitution concentrations in the range of $x \leq 0.2$, the Se substitution leads to the disappearance of the CDW-like state. Besides, a slight increase of $T_c$ can be seen with the highest $T_c$ of about 2.83 K for the optimal doping $x$ = 0.10. However, further augmentation of the dopant content ($x \geq 0.25$) caused the reappearance of noticeable anomalies in both resistivity and magnetization, indicating the CDW-like transition retakes place, which is highly related to the disorder effects created by selenium doping. Nevertheless, these findings call for further studies to determine the possible explanations behind this kind of behavior.

## Acknowledgments

This work is supported by the National Natural Science Foundation of China (Grants No. 11922415), Guangdong Basic and Applied Basic Research Foundation (2019A1515011718), the Fundamental Research Funds for the Central Universities (19lgzd03), Key Research & Development Program of Guangdong Province, China (2019B110209003), and the Pearl River Scholarship Program of Guangdong Province Universities and Colleges (20191001). The work at IOPCAS is supported by the NSFC (12025408, 11921004, 11904391), the National Key R&D Program of China (2018YFA0305702). Y. Y. P. is grateful for financial support from the National Natural Science Foundation of China (Grant No. 11974029).

**Figures caption:**

**Figure 1. (a)** Rietveld refinement for the representative $CuIr_2Te_{3.9}Se_{0.1}$. **(b)** Powder X-ray diffraction patterns for $CuIr_2Te_{4-x}Se_x$ ($0 \leq x \leq 0.5$). **(c)** The dependence of unit-cell constants $a$ and $c$ on the doping concentration $x$.

**Figure 2** Electrical transport and magnetic properties of $CuIr_2Te_{4-x}Se_x$. **(a)** The temperature-dependent resistivity for the polycrystalline $CuIr_2Te_{4-x}Se_x$. **(b)** The magnified view of the normalized resistivity ($\rho/\rho_{300K}$) at the superconducting transition range for the polycrystalline $CuIr_2Te_{4-x}Se_x$ at low temperatures. **(c)** Magnetic susceptibilities for $CuIr_2Te_{4-x}Se_x$ ($0 \leq x \leq 0.5$) near the superconducting transitions region measured under 10 Oe applied field. **(d)** Magnetization curves measured under 10 kOe for $CuIr_2Te_{4-x}Se_x$ ($0.3 \leq x \leq 0.5$).

**Figure 3** Temperature-dependent PXRD patterns for the representative samples: (a) $CuIr_2Te_4$ and (b) $CuIr_2Te_{3.5}Se_{0.5}$ at 20 K, 100 K and 300 K.

**Figure 4 (a)** Low-temperature $C_p/T$ at zero (solid green circles) and 10 kOe field (solid blue circles) as a function of $T^2$, the inset shows $C_{el}/T$ vs $T$. **(b)** The lower critical field for $CuIr_2Te_{3.9}Se_{0.1}$ with the fitting lines using the equation $\mu_0H_{c1}(T) = \mu_0H_{c1}(0)(1-\left(\frac{T}{T_c}\right)^2)$. The bottom and upper insets illustrate the magnetic susceptibilities $M(H)$ curves and $M-M_{Fit}(H)$ at various temperatures, respectively. **(c-d)** The upper critical fields for $CuIr_2Te_{3.9}Se_{0.1}$ and $CuIr_2Te_{3.8}Se_{0.2}$. The data are extracted from $\rho(T,H)$ and fitted by GL and WHH models for different criteria. The insets depict the temperature-dependent resistivity under different magnetic field $\rho(T,H)$ curves for $CuIr_2Te_{3.9}Se_{0.1}$ and $CuIr_2Te_{3.8}Se_{0.2}$, respectively.

**Figure 5** The phase diagram of $T_c$ and $T_{CDW}$ versus Se doping content. The $T_c$ data has been extracted from $\rho/\rho_{300K}$ (T) and $\chi(T)$. $T_{CDW}$ has been extracted from the cooling data of the zero-field $\rho/\rho_{300K}$ (T) and $\chi(T)$ under 10 kOe.

**Table 1.** Se content ($x$) dependence of the residual resistance ratio (RRR = $R_{300K}/R_{8K}$), superconducting transition temperature ($T_c$), and CDW transition temperature ($T_{CDW}$).

**Table 2** Superconducting parameters of different ternary telluride chalcogenides compounds.

**Figure 1**

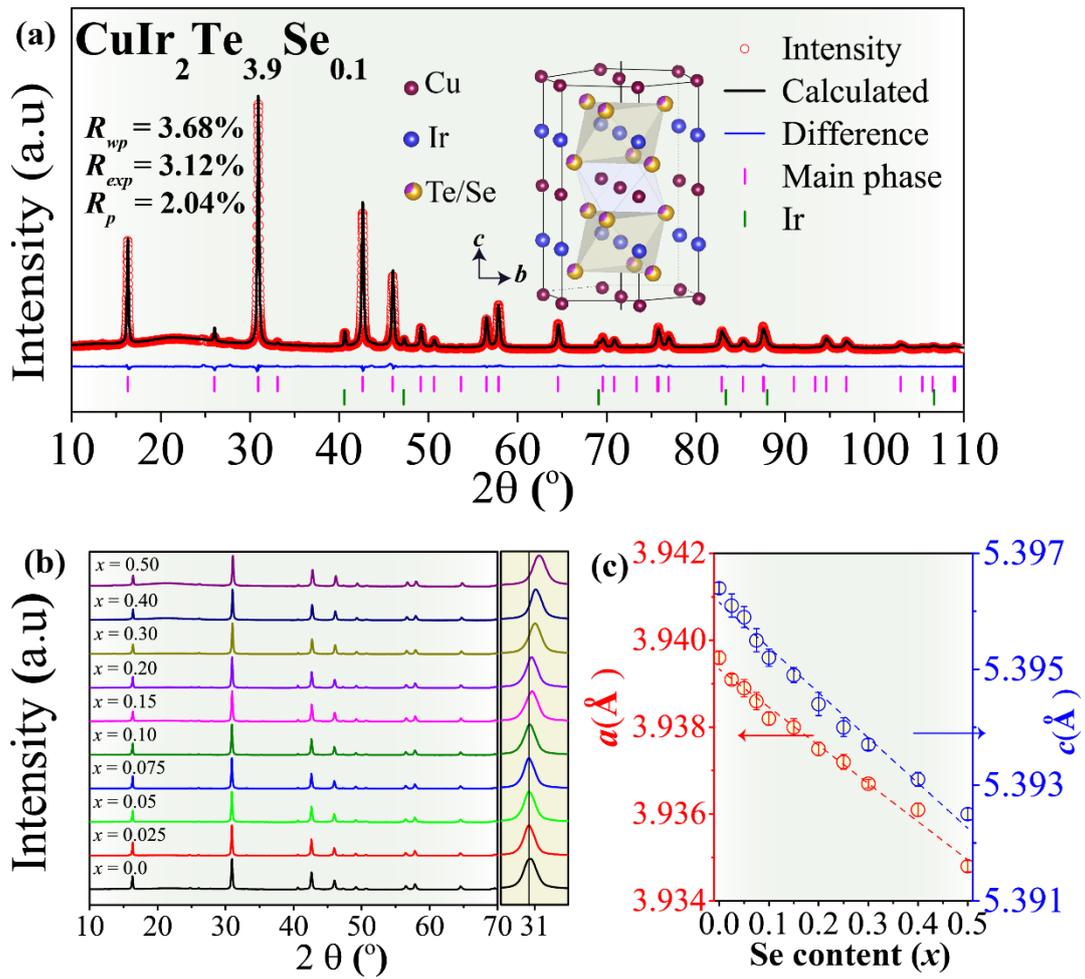

**Figure 2**

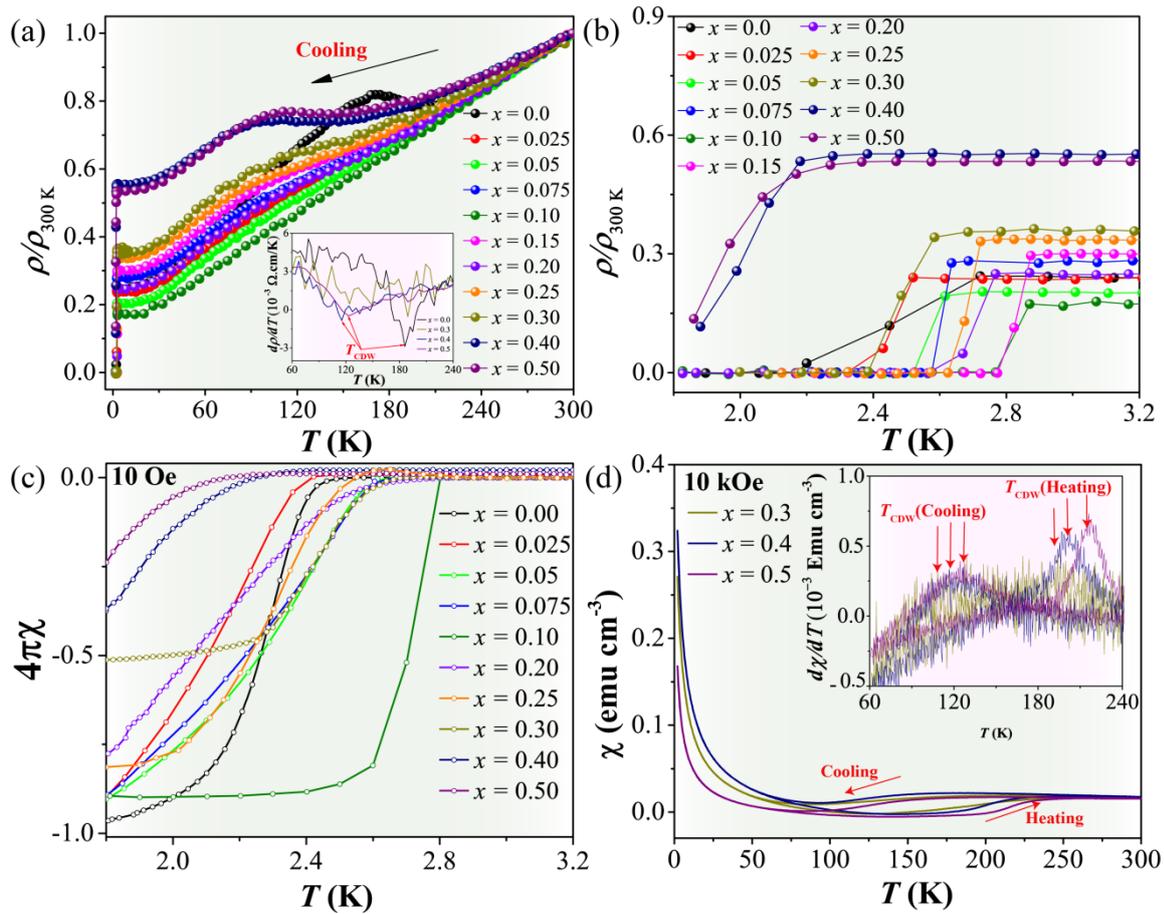

**Figure 3**

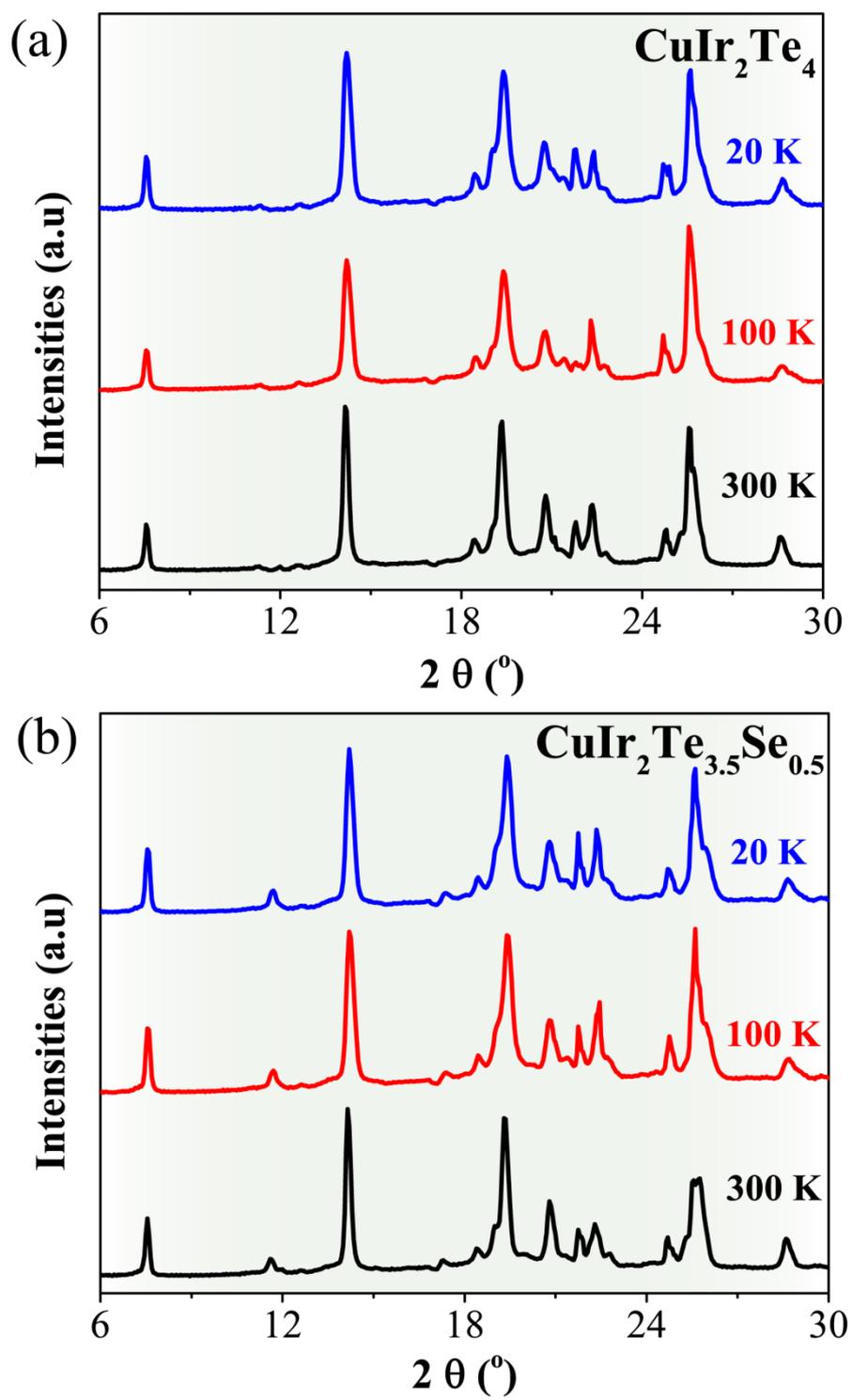

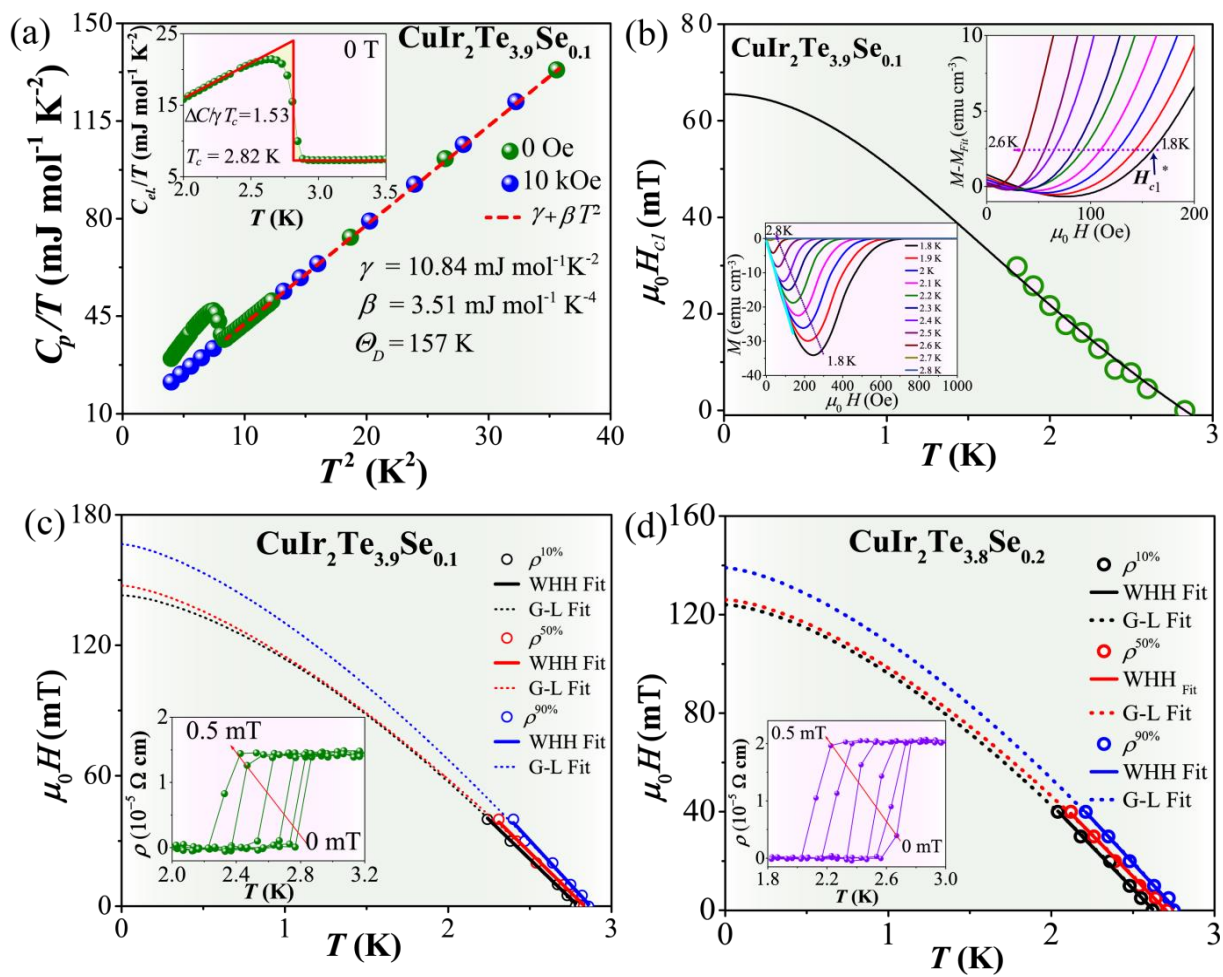

**Figure 5**

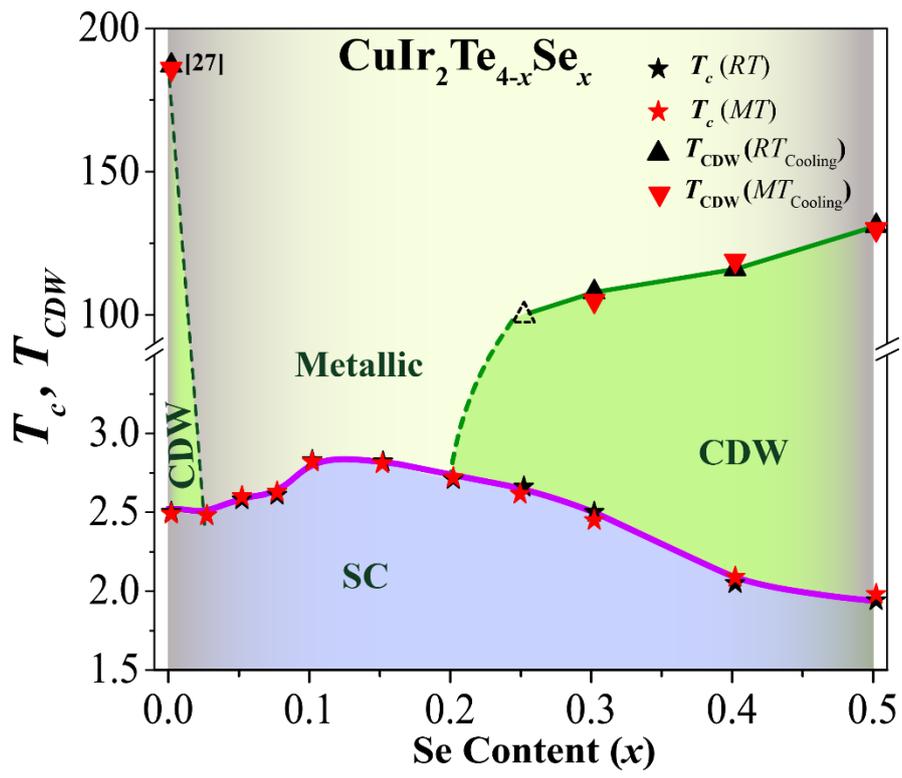

Table 1

| Se content ($x$) | RRR = $R_{300K}/R_{8K}$ | $T_c$ (K) | $T_{CDW}$ (K) |
|---|---|---|---|
| 0 | 4.16 | 2.50 | 187 |
| 0.025 | 4.34 | 2.48 | |
| 0.05 | 4.87 | 2.58 | |
| 0.075 | 3.70 | 2.61 | |
| 0.1 | 5.88 | 2.83 | |
| 0.15 | 3.49 | 2.82 | |
| 0.2 | 4.16 | 2.71 | |
| 0.25 | 3.02 | 2.68 | 100 |
| 0.3 | 2.71 | 2.50 | 108 |
| 0.4 | 1.81 | 2.05 | 116 |
| 0.5 | 1.87 | 1.94 | 131 |

**Table 2**

| Parameter \ Material | CuIr$_2$Te$_{3.9}$Se$_{0.1}$ (This work) | CuIr$_2$Te$_{3.8}$Se$_{0.2}$ (This work) | CuIr$_2$Te$_4$ [33] | CuIr$_{1.95}$Ru$_{0.05}$Te$_4$ [58] | CuIr$_2$Te$_{3.9}$I$_{0.1}$ [57] | Cu$_{0.25}$Zn$_{0.25}$IrTe$_2$ [59] |
|---|---|---|---|---|---|---|
| $T_c$ (K) | 2.83 | 2.71 | 2.5 | 2.79 | 2.95 | 2.82 |
| $\gamma$ (mJ mol$^{-1}$ K$^{-2}$) | 10.84 | | 12.05 | 12.26 | 12.97 | 13.37 |
| $\beta$ (mJ mol$^{-1}$ K$^{-4}$) | 3.51 | | 1.97 | 1.87 | 3.03 | 1.96 |
| $\Theta_D$ (K) | 157 | | 190 | 193 | 165 | 190.6 |
| $\Delta C/\gamma T_c$ | 1.51 | | 1.5 | 1.51 | 1.46 | 1.45 |
| $\lambda_{ep}$ | 0.65 | | 0.63 | 0.65 | 0.70 | 0.66 |
| $N(E_F)$ (states/eV/f.u.) | 3.11 | | 3.1 | 3.15 | 3.24 | 3.41 |
| $\mu_0 H_{c1}(0)$ (mT) | 66 | | 28 | 98 | 24 | 62 |
| $\mu_0 H_{c2}(0)$ (mT) ($\rho_{N50\%}$ G-L theory) | 148 | 127 | 145 | | 232 | 198 |
| $\mu_0 H_{c2}(0)$ (mT) ($\rho_{N50\%}$ WHH theory) | 144 | 125 | 120 | 247 | 188 | |
| $-dH_{c2}/dT_c$ (mT/K) | 73.3 | 66 | 66 | 125 | | |
| $\mu_0 H^P$ (T) | 5.26 | 5.04 | 4.65 | 5.24 | 5.49 | 5.26 |
| $\xi_{GL}$ (nm) | 47.18 | 51.34 | 52.8 | 36.3 | 41.9 | 40.7 |